\documentclass[a4paper,twocolumn,11pt,allowtoday,accepted=2026-03-22]{quantumarticle}
 \pdfoutput=1
\usepackage{graphicx,color}
\usepackage{float}
\usepackage{amsmath,amssymb,amsfonts}
\usepackage{bm}
\usepackage[pdftex,colorlinks=true, linkcolor = blue, citecolor=blue,urlcolor=blue, bookmarksnumbered=true, bookmarksopen=true]{hyperref}
\usepackage[normalem]{ulem}
\usepackage[numbers]{natbib}
\usepackage{braket}
\usepackage{algorithm}
\usepackage{algpseudocode}
\makeatletter
\def\BState{\State\hskip-\ALG@thistlm}
\makeatother

\begin{document}

\title{Recursive algorithm for constructing antisymmetric fermionic states in first quantization mapping}

\author{E. Rule}
\affiliation{
  Theoretical Division, Los Alamos National Laboratory, Los Alamos, New Mexico 87545, USA}
\affiliation{Department of Physics, University of California, Berkeley, CA 94720, USA}
\author{I. A. Chernyshev}
\affiliation{
  Theoretical Division, Los Alamos National Laboratory, Los Alamos, New Mexico 87545, USA}
  \author{I. Stetcu}
\affiliation{
  Theoretical Division, Los Alamos National Laboratory, Los Alamos, New Mexico 87545, USA}
\author{J. Carlson}
\affiliation{
  Theoretical Division, Los Alamos National Laboratory, Los Alamos, New Mexico 87545, USA}

\author{R. Weiss}
\affiliation{
Department of Physics, Washington University in St. Louis, St. Louis, Missouri 63130, USA}
  
\date{March 23, 2026}

\begin{abstract}
We devise a deterministic quantum algorithm to produce antisymmetric states of single-particle orbitals in the first quantization mapping. Unlike sorting-based antisymmetrization algorithms, which require ordered input states and high Clifford-gate overhead, our approach initializes the state of each particle independently. For a system of $\eta$ particles and $N$ single-particle states, our algorithm prepares antisymmetrized states of non-trivial localized (\textit{e.g.,} Hartree-Fock) orbitals using $O(\eta^2\sqrt{N})$ $T$-gates, outperforming alternative algorithms when $\eta\lesssim \sqrt{N}$. To achieve such scaling, we require $O(\sqrt{N})$ dirty ancilla qubits for intermediate calculations. Knowledge of the single-particle states to be antisymmetrized can be leveraged to further improve the efficiency of the circuit, and a measurement-based variant reduces gate cost by roughly a factor of two. We show example circuits for two- and three-particle systems and discuss the generalization to an arbitrary number of particles. For a specific three-particle example, we decompose the circuit into Clifford$+T$ gates and study the impact of noise on the prepared state.

\end{abstract}
\maketitle

\section{Introduction}

The dynamics of quantum few- and many-body systems is often described by unitary
time evolution
\[
    |\Psi(t)\rangle=e^{-iHt}|\Psi(0)\rangle
\]
of a specific initial state $| \Psi(0) \rangle$ with Hamiltonian $H$ typically consisting of two- and possibly three-particle interactions. An exact solution for the time evolution is usually prohibitive on classical computers because
of the large number of many-body states involved, even if there are a modest number of particles.

For specific types of problems, approximate solutions have been developed: At large energies and
momentum transfers, semiclassical methods are often employed \cite{FORD1959259,brack2003semiclassical,brink_2025_cd47w-rxr81}. For low energies and large numbers of particles, the time-dependent density functional theory has been successfully used to describe various phenomena \cite{LaurentJacquemin2013_TDDFTBenchmarksReview,PhysRevLett.116.122504}. Even in these limiting cases, in nuclear physics, the energy density functional cannot be exactly derived from first principles and phenomenology must be involved. For few nucleons, the situation is even more dire, as mean-field approaches usually work less well in these cases. 

Quantum computing can in principle be used to directly simulate the full quantum evolution by
encoding the large number of states in a relatively modest set of qubits. The relevant Hamiltonians typically consist of two- and possibly three-particle interactions (often Coulomb for condensed-matter systems \cite{chemla_2001}, effective potentials for
quantum liquids like Helium \cite{1973AuJPh..26...43K}, or nucleon-nucleon interactions for nuclear physics \cite{Piarulli:2019cqu}). In any case, the full Hamiltonian is symmetric under the exchange of particles, and the physically relevant solutions are either symmetric for identical bosons or antisymmetric for identical fermions.

Different mappings have been devised for encoding many-body fermionic systems on quantum hardware. 
They have distinct advantages and disadvantages, and the best approach depends on the particular problem under consideration. Second-quantization mappings \cite{AbramsLloyd1997,PhysRevA.64.022319,2005Sci309.1704A} are in general simpler to use for fermionic systems, as the encoded many-body states are antisymmetric by default. In this approach, however, the number of required qubits scales \textit{linearly} with the number of single-particle states, leading to inefficiencies when the number of single-particle states is much larger than the number of particles. This is often the case when a high-resolution result is required for Hamiltonians with short-range interactions or high-momentum scattering. In simulations of chemical or nuclear reactions, which require asymptotic solutions to extract information like cross sections, the number of logical qubits necessary to describe such systems is well beyond the capability of near-term quantum devices. In this case, the first quantization mapping \cite{Lloyd:1996aai,AbramsLloyd1997,firstQuantization}, where the number of qubits scales \textit{logarithmically} with the number of single-particle states and linearly with the number of particles, is more efficient and has more favorable prospects to be implemented on upcoming devices. 

Under permutation of particle labels, physical states of fermionic systems must transform with the correct antisymmetric phase ($+1$ for even permutations and $-1$ for odd permutations). In the first-quantization mapping, however, many-body fermionic states are not antisymmetrized a priori. Commonly, one instead begins with a product state of orthogonal single-particle orbitals. Assuming that the Hamiltonian consists of one-body terms plus higher-body interactions that are expressible in terms of auxiliary fields, then unitary time evolution will result in a sum of product states of orthogonal single-particle orbitals. In such scenario, depending on the physical observables of interest, it is not strictly necessary to completely antisymmetrize the initial wave function. Instead, one can obtain expectation values of properly-antisymmetrized many-body operators by summing over all $k$-tuples of particles that can interact through a $k$-body operator, with the proper signs under particle exchanges. For a system of $\eta$ particles, this requires evaluation of $^\eta C_k$ $k$-body operators (or sampling from them) in order to recover the expectation value of a $k$-body operator. If instead one fully antisymmetrizes the wave function, one can exploit this antisymmetry to evaluate the $k$-body operator only once on, say, the first $k$ particles with an appropriate normalization factor. This is usually the more efficient approach.

One can obtain such an antisymmetric state either by a deterministic construction or by projecting from an arbitrary initial wave function onto the antisymmetric component. The algorithm we pursue in this work performs a deterministic construction of the antisymmetric state, but we briefly outline here the projection technique, as it may have some specialized use cases.

Projection onto the antisymmetric component can be accomplished by (1) adding an infinite potential for any doubly-occupied states and (2) adding a two-body permutation term to the Hamiltonian. The infinite potential in (1) does not impact fermionic states. Since the Hamiltonian is symmetric, remaining states without double occupancy can be separated by adding a permutation term to the Hamiltonian:  $H_\mathrm{perm} = \sum_{i<j} P_{ij} $, where $P_{ij}$ permutes all coordinates between particles $i$ and $j$. $H_\mathrm{perm} $ commutes with the original Hamiltonian, and so we can simultaneously diagonalize the original Hamiltonian $H$ and $H_\mathrm{perm}$ to obtain eigenstates of good exchange symmetry. 

For $\eta$ particles, the eigenvalues of $H_\mathrm{perm}$ are integer-valued, ranging from $-\eta(\eta-1)/2$ for fermionic many-body states to $\eta(\eta-1)/2$ for bosonic states. Intermediate eigenvalues correspond to mixed-symmetry states. Such a spectrum of constantly-spaced eigenvalues can be easily projected using standard quantum phase estimation techniques \cite{NielsenMichaelA.2000Qcaq,Stetcu-2023proj}. By including an appropriate coupling to $H_\mathrm{perm}$ to guarantee that bosonic and mixed symmetry states are higher in energy than fermionic states, one can still use variational techniques to create initial states. The additional permutation Hamiltonian contains $O(\eta^2)$ terms, the same as the scaling of typical two-body potentials.

However, except for systems with very few particles, a specific product state has very low overlap with the corresponding antisymmetric state. Therefore, this projection technique is mostly useful for either detecting symmetry violations introduced by noise in the quantum hardware or for obtaining a fully antisymmetric wave function from a state with a large antisymmetric component. The latter could arise, for example, in scattering calculations where one antisymmetrizes the target many-body system but does not antisymmetrize the projectile with the target fermions.

In general, dynamical problems are more efficiently treated by a full antisymmetrization of the initial state. The efficient mapping introduced in Ref. \cite{PhysRevResearch.4.023154}, where fully antisymmetric basis states are mapped in combinations of up and down qubit states, can, in principle, be an alternative. In this encoding, the number of qubits necessary to solve a many-body problem scales logarithmically with the number of antisymmetric \textit{many-body} states. However, this approach, while the most efficient in terms of qubit requirements for the same many-body problem, has the disadvantage that the operator mapping (including the Hamiltonian) will generically result in a large number of Pauli strings \cite{PhysRevX.8.011044,Li:2024}.

Early on, Abrams \& LLoyd \cite{AbramsLloyd1997} devised a quantum algorithm capable of antisymmetrizing a particular class of input states: linear superpositions of \textit{ordered} basis states of the form $\ket{r_1\ldots r_\eta}$ where $r_1<\ldots < r_\eta$ are integers that label the single-particle states. More recently, Berry \textit{et al}. \cite{Berry2018a} introduced an algorithm with improved gate depth and complexity, but with the same requirement of ordered input states. Subsequent work \cite{PhysRevA.106.032428,Babbush:2023sas,PRXQuantum.6.020319} demonstrated that --- despite this restriction on the form of the input states --- this antisymmetrization algorithm is compatible with state-preparation procedures that together can be used to prepare one (or more) Slater determinants composed of single-particle orbitals, each with its own degree of complexity. 

In this paper, we present an iterative algorithm to construct the antisymmetric state of $\eta$ particles by successively building antisymmetric states of 2, 3, $\ldots$ up to $\eta$ particles. Our method does not place any restriction on the form of the states to be antisymmetrized (other than the requirement that they be orthogonal) and performs simultaneously the antisymmetrization and state preparation necessary to prepare Slater determinants of complicated single-particle orbitals. 

In addition to the physical qubits required to represent the state of each particle, our method requires $\eta-1$ ancillae to manipulate the state. At the end of each iterative step, the ancillae are disentangled from the physical qubits and no measurements are required. Nonetheless, a measurement-based variant of our approach that preserves 100\% of the success probability can reduce the complexity of the quantum circuit, as we describe below. 

The paper is organized as follows: In Sec. \ref{sec:algorithm}, we present the general algorithm, describe the measurement-based variant, and discuss the explicit application of these approaches to two- and three-particle systems. In Sec. \ref{sec:other_algorithms}, we study the scaling with the number of particles and the number of states per particle, and compare our approach against existing antisymmetrization algorithms. In Sec. \ref{sec:examples}, we show a concrete example for three particles, expand the circuit in Clifford+$T$ gates, and estimate the noise level that would still allow us to obtain good antisymmetrization properties of the prepared wave function. Finally, in Sec. \ref{sec:conclusions} we present a summary and our conclusions. To aid the reader, we include a pair of Appendices containing further details of our circuit implementations.

\begin{figure*}
    \centering
\includegraphics[scale=1.0]{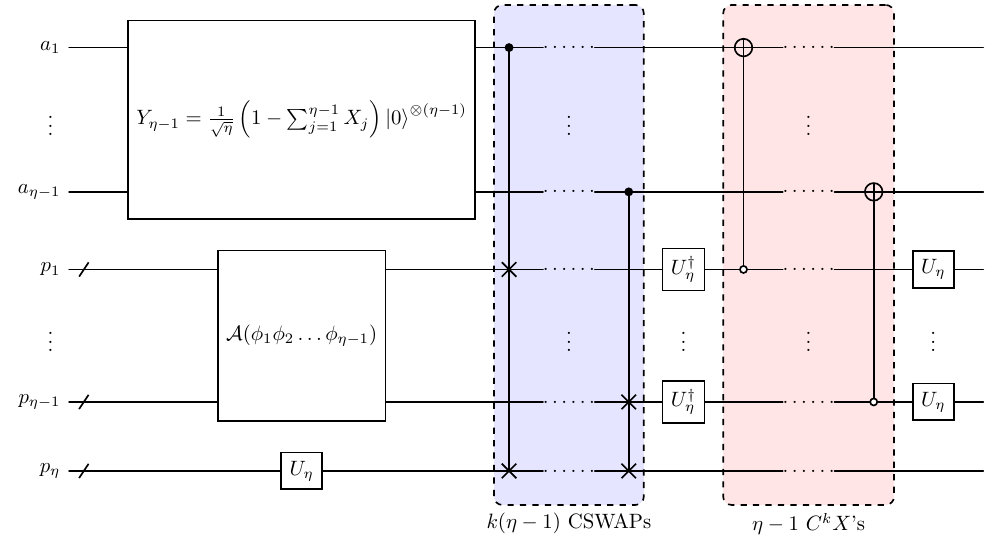}
    \caption{Quantum circuit that prepares the antisymmetric state on $\eta$ particles from the antisymmetric state on $\eta-1$ particles. $\eta-1$ ancilla qubits are prepared in the $Y_{\eta-1}$ state (see Appendix \ref{app:ancilla_prep}). Controlling on the state of the ancillae, we perform \textsc{swap} operations between each of the previous $\eta-1$ particles and the $\eta$-th particle (blue shaded region). To uncompute the ancillae, we detect if a \textsc{swap} occurred by acting the operator $U_\eta^\dagger$ on each of the $\eta-1$ previous particles and controlling on the particle being in the state $\ket{0}^{\otimes k}$. The multi-controlled $X$-gates that uncompute the ancilla qubits (red shaded region) can be performed in parallel. Finally, we restore the state of the first $\eta-1$ particles by acting $U_\eta$ on each.}
    \label{fig:asymN}
\end{figure*}

\section{Description of the algorithm}
\label{sec:algorithm}

In this work, we distinguish between single-particle basis states and single-particle states, often called orbitals (such as Hartree-Fock or Kohn-Sham orbitals). The latter are typically linear combinations of the former and may break underlying symmetries of the Hamiltonian. Nonetheless, a single Slater determinant constructed from these orbitals serves as a useful antisymmetric initial state. Although approximate and generally not sufficient to obtain an exact eigenvector through variational techniques, it generally exhibits substantial overlap with the targeted low-lying eigenstates and can be improved through variational optimization that preserves the antisymmetrization, followed by projection onto the desired eigenstate. (For weakly correlated systems, the Hartree-Fock approximation is an excellent starting point and, depending on the targeted error, would probably not require subsequent projection.) In first quantization, the single-particle basis states are assumed to correspond to integers.

We start in Sec. \ref{sec:gen_circ} with a circuit that is more applicable to current hardware, if mid-circuit measurements are not available or limited.  Although the circuit is general and would work for arbitrary orthogonal orbitals, in Sec. \ref{sec:improved} we introduce a measurement-based approach that reduces the depth of the circuit, and discuss several other optimization strategies. 

\subsection{General circuit}
\label{sec:gen_circ}
Given $N$ single-particle basis states, they can be represented as a superposition of $k=\lceil\log_2N\rceil$ qubits, and the many-body state of $\eta$ particles can thus be encoded on a total of $\eta k$ qubits. We assume that we have a set of orthogonal single-particle wave functions $\phi_n$, with $n=1,2,\ldots, \eta$, and the associated unitary operators (quantum circuits) $U_n$ that prepare those states on $k$ qubits from state $\ket{0}^{\otimes k}$:
\begin{equation}
    \ket{\phi_n}=U_n \ket{0}^{\otimes k}.
\end{equation}
We also assume that we can prepare the inverse of the quantum circuit $U_n$, denoted by $U_n^\dagger$, by reversing the order of the gates, so that when applied to the circuit producing $\ket{\phi_n}$, we obtain $\ket{0}^{\otimes k}$. The orthogonality of the single-particle wave functions implies that for $m\neq n$
\begin{equation}
    \langle \phi_m|\phi_n\rangle=\langle0|^{\otimes k} U_m^\dagger U_n |0\rangle^{\otimes k}=0,
\end{equation}
or, in other words, $U_m^\dagger U_n\ket{0}^{\otimes k}$ does not contain the state $\ket{0}^{\otimes k}$.

Our algorithm is recursive: we progressively construct the 2-, 3-, up to $\eta$-particle antisymmetric states, each time using the previous results as a new starting point. The general algorithm is described below in quantum pseudocode as Algorithm \ref{algo:recur_antisym}, and the corresponding quantum circuit is depicted in Fig. \ref{fig:asymN}. Assume that we already have produced the antisymmetric state for $\eta-1$ particles $\mathcal{A}(\phi_1\ldots\phi_{\eta-1})$ on the first $(\eta-1)k$ qubits and the state $\phi_\eta$ on the $k$ qubits associated with particle $\eta$. In addition, we will require $\eta-1$ ancilla qubits, which will be prepared in the state 
\begin{equation}
    \ket{Y_{\eta-1}}\equiv\frac{1}{\sqrt{\eta}}\left(1-\sum_{j=1}^{\eta-1}X_j\right)\ket{0}^{\otimes (\eta-1)},
\end{equation}
consisting of an equal superposition of the zero state and all states with a single bit flip. Details on the preparation of the $Y_\eta$ state are provided in Appendix \ref{app:ancilla_prep}. In each term, the location of the flipped bit encodes which (if any) of the previous $\eta-1$ particles will be swapped with the $\eta$-th particle. The relative minus sign ensures the correct antisymmetric behavior under particle exchange.

Controlling on ancilla $i$, we swap the qubit states associated to particle $\eta$ with those of particle $i$ for each $1\leq i< \eta$. After performing a total of $(\eta-1)k$  controlled-swaps (\textsc{cswap}s), we have prepared the qubits in the antisymmetric state on $\eta$ particles, but they are entangled with the ancillae. We begin to uncompute the ancillae by applying $U_\eta^\dagger$ to the first $\eta-1$ particle registers. Because we construct the antisymmetric state recursively, particle $i<\eta$ can only be in state $\phi_\eta$ if a swap occurred. Thus, after applying $U_\eta^\dagger$, particle $i<\eta$ will be in state $\ket{0}^{\otimes k}$ if a swap occurred and some state not containing $\ket{0}^{\otimes k}$ if no swap occurred. We then uncompute ancilla $i$ by performing a multi-controlled $C^{k}X$, controlled on the qubits associated to particle $i<\eta$ being in state $\ket{0}^{\otimes k}$ and targeting ancilla $i$. As each of the $\eta-1$ particles is associated to a unique ancilla qubit, these operations can be performed in parallel. Finally, the antisymmetric state on $\eta$ particles is restored by applying $U_\eta$ to the first $\eta-1$ particle registers. 

\begin{algorithm}[H]
\caption{Recursive Antisymmetrization}\label{algo:recur_antisym}
\begin{algorithmic}[1]
\State \textbf{Input: $\mathcal{A}(\phi_1\ldots\phi_{\eta-1})$, $U_\eta$}
\State \textbf{Output: $\mathcal{A}(\phi_1\ldots\phi_\eta)$}
\State $(p_1,\ldots,p_{\eta-1}) \gets \mathcal{A}(\phi_1\ldots\phi_{\eta-1})$
\State $p_\eta\gets U_\eta\ket{0}^{\otimes k}$
\State $(a_0,\ldots,a_{\eta-1})\gets Y_{\eta-1}\equiv\frac{1}{\sqrt{\eta}}\left(1-\sum_{j=1}^{\eta-1}X_j\right)\ket{0}^{\otimes (\eta-1)}$
\For{$i =1,\ldots,\eta-1$} 
\State Swap $p_i$ and $p_\eta$ controlled on $a_i$ being in state $\ket{1}$
\EndFor
\For{$i =1,\ldots,\eta-1$} 
\State $p_i\gets U^\dagger_\eta p_i$
\State Map $a_i$ to $\ket{0}$ controlled on $p_i$ being in state $\ket{0}^{\otimes k}$
\State $p_i\gets U_\eta p_i$
\EndFor
\State \textbf{return} $\mathcal{A}(\phi_1\ldots\phi_\eta)$
\end{algorithmic}
\end{algorithm}

\begin{figure}
    \centering
\includegraphics[scale=0.9]{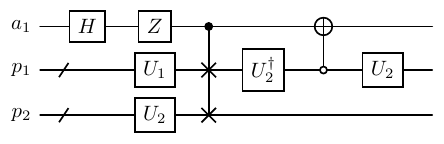}
    \caption{Quantum circuit for generating the antisymmetric state of two particles in single-particle states.}
    \label{fig:asym2}
\end{figure}

We will now discuss explicit examples for two and three particles: The circuit for the antisymmetrization of two particles in states $\phi_1$ and $\phi_2$ following the algorithm presented above is shown in Fig. \ref{fig:asym2}. In this case, the ancilla state $Y_1=(\ket{0}-\ket{1})/\sqrt{2}$ can be easily prepared with a Hadamard followed by a $Z$-gate. After the controlled-swap application, the circuit encodes the state $\frac{1}{\sqrt{2}}(\phi_1\phi_2\ket{0}_a-\phi_2\phi_1\ket{1}_a)$. Applying the reversed circuit $U_2^\dagger$ reduces the particle state to $\ket{0}^{\otimes k}$ if particle $p_1$ is in the state $\phi_2$, and some combination of qubit states that does not include $\ket{0}^{\otimes k}$ if the particle is in the state $\phi_1$. The multi-controlled \textsc{not} operation uncomputes the ancilla. The final application of $U_2$ restores the state of particle $p_1$ to either $\phi_1$ or $\phi_2$, as appropriate. Obviously, we could have uncomputed the ancilla by instead applying $U_1^\dagger$ to particle $p_2$, but this approach does not generalize to more than two particles. 

\begin{figure*}
    \centering
\includegraphics[scale=0.95]{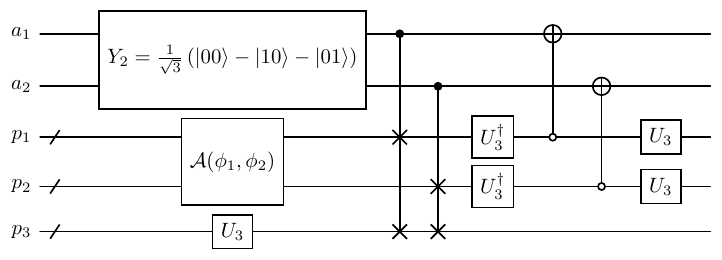}
     \caption{Same as in Fig. \ref{fig:asymN}, but for three particles starting with the previously computed antisymmetric state of two particles (see Fig. \ref{fig:asym2}). The swaps are controlled on either ancilla $a_1$ or $a_2$ being in state $\ket{1}$, depending on whether the swap involves particle $p_1$ or $p_2$, respectively.}
    \label{fig:asym3}
\end{figure*}
 
In order to prepare the three-body antisymmetric state, we use the results for two particles as input to the circuit shown in Fig. \ref{fig:asym3}. The derivation then proceeds in a similar manner. After performing two controlled swaps, the three-particle wave function is encoded as 
\begin{equation*}
\begin{split}
 \frac{1}{\sqrt{3!}}[ &(\phi_1\phi_2 \phi_3 - \phi_2\phi_1 \phi_3)\ket{00}_a \\
    - &(\phi_3\phi_2\phi_1-\phi_3\phi_1\phi_2) \ket{10}_a \\
    - &(\phi_1\phi_3\phi_2-\phi_2\phi_3\phi_1)\ket{01}_a]. 
    \end{split}
\end{equation*}
Similarly to the two-body case, we now apply $U_3^\dagger$ to particles $p_1$ and $p_2$, reducing the state $\phi_3$ to $\ket{0}^{\otimes k}$, wherever present. This operation is then followed by the uncomputing of the two ancilla qubits with two multi-controlled operations. Finally, the state of particles $p_1$ and $p_2$ is restored by applying $U_3$. For systems of more than three particles, the algorithm proceeds further in the same manner, following Fig. \ref{fig:asymN}. 

To prepare the $\eta$-particle antisymmetric state, each $U_n$ is applied once to prepare particle $n$ in state $\phi_n$. In addition, at the $i$-th iterative step, $U_{i+1}$ and $U_{i+1}^\dagger$ are each applied $i$ times for the purpose of uncomputing the ancillae. Thus, we require a total of $\eta + \eta(\eta-1)/2=\eta(\eta+1)/2$ applications of $U_n$ operators and $\eta(\eta-1)/2$ applications of $U_n^\dagger$ operators. This approach to antisymmetrization is most efficient when the state-preparation unitaries $U_n$ and $U_n^\dagger$ are simple quantum circuits; in particular, if the single-particle states that are to be antisymmetrized correspond to a product state of $k$ qubits. For example, if each state is a binary representation of an integer, then each $U_n$ is simply the product of Pauli $X$-gates that produces the corresponding binary representation.

In principle, the structure of our algorithm allows one to exploit information about the single-particle states to simplify the resulting circuit. Over the course of the algorithm, the state-preparation unitary $U_{n}$ is applied $n$ times and $U_{n}^\dagger$ is applied $n-1$ times. Thus, if one of the single-particle states is significantly more costly to prepare or more sensitive to noise than the other $\eta-1$ states, it makes sense to label this state as $\phi_1$ in order to minimize the associated state-preparation costs. Likewise, the simplest state to prepare should be added in the last iterative step. Although $U_n$ can act on all $k$ qubits associated to particle $n$, in practice this may not be the case. For example, if the initial physical state were to consist of two molecules or two nuclei prepared on opposite sides of a finite lattice volume, then the state of the nucleons on either side could in principle be prepared using roughly half of the available physical qubits.

\subsection{Measurement-based variant}
\label{sec:improved}

\begin{figure}
    \centering
    \includegraphics[scale=0.87]{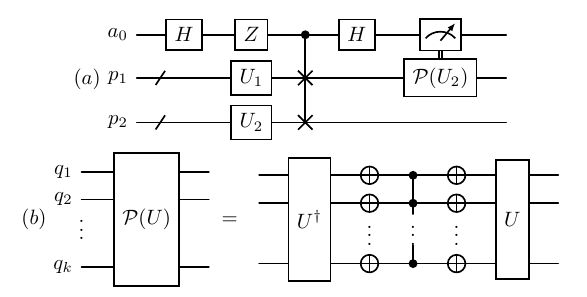}
    \caption{Measurement-based antisymmetrization of two particles: (a) general circuit, which requires a phase gate in case the ancilla $a_0$ is measured in state $\ket{1}$. The requisite phase gate $\mathcal{P}(U)$ is defined in panel (b).}
    \label{fig:asym2-meas}
\end{figure}

The circuit shown in Fig. \ref{fig:asymN} is general, but for orbitals that are superpositions of many basis states the circuit cost becomes dominated by the $U_n$ and $U_n^\dagger$ operators, which together must be applied a total of $\eta^2$ times to produce the $\eta$-particle antisymmetric state. Additionally, disentangling the ancillae requires additional multi-controlled operations. We can reduce the number of applications of $U_n$ and $U_n^\dagger$ and eliminate some number of multi-controlled gates by adopting a measurement strategy. Figure \ref{fig:asym2-meas}(a) illustrates the resulting algorithm for antisymmetrizing two particles, which now includes an application of a Hadamard gate on the ancilla and a measurement. If the measurement results in state $\ket{0}$, the prepared state is antisymmetric, and no further operation is necessary. If the ancilla is measured in state $\ket{1}$, then the physical state is in the symmetric combination $(\phi_1\phi_2+\phi_2\phi_1)/\sqrt{2}$. To correct this, we apply to particle $p_1$ the operator $U_2^\dagger$, followed by a multi-controlled $Z$ operation to restore the antisymmetry, and finally $U_2$ to restore the state $\phi_2$, as shown in Fig. \ref{fig:asym2-meas}(b). The ancilla can then be reset and reused, if desired. As discussed earlier, one can leverage information about the single-particle state to reduce the number of gates, if appropriate.

The same approach can be generalized to prepare the antisymmetric state of $\eta$ particles by applying Hadamard gates to all $\eta-1$ ancilla qubits, followed by conditional phase operations on the relevant particle states based on the measurement outcomes of the ancilla qubits. Thus, we introduce Algorithm \ref{algo:measure_antisym}, described in quantum pseudocode as follows:

\begin{algorithm}[H]
\caption{Measurement-based Antisymmetrization}\label{algo:measure_antisym}
\begin{algorithmic}[1]
\State \textbf{Input: $\mathcal{A}(\phi_1\ldots\phi_{\eta-1})$, $U_\eta$}
\State \textbf{Output: $\mathcal{A}(\phi_1\ldots\phi_\eta)$}
\State $(p_1,\ldots,p_{\eta-1}) \gets \mathcal{A}(\phi_1\ldots\phi_{\eta-1})$
\State $p_\eta\gets U_\eta\ket{0}^{\otimes k}$
\State $(a_0,\ldots,a_{\eta-1})\gets Y_{\eta-1}\equiv\frac{1}{\sqrt{\eta}}\left(1-\sum_{j=1}^{\eta-1}X_j\right)\ket{0}^{\otimes (\eta-1)}$
\For{$i =1,\ldots,\eta-1$} 
\State Swap $p_i$ and $p_\eta$ controlled on $a_i$ being in state $\ket{1}$
\EndFor
\For{$i =1,\ldots,\eta-1$} 
\State $a_i\gets Ha_i$
\State Measure $a_i\rightarrow c_i$
\EndFor
\If{$\sum_jc_j \leq \lfloor \eta/2\rfloor$}
\For{$i=1,\ldots,\eta-1$}
\If{$c_i=1$}
\State $p_i\gets \mathcal{P}(U_\eta)p_i$
\EndIf
\EndFor
\EndIf
\If{$\sum_jc_j> \lfloor \eta/2\rfloor$}
\For{$i=1,\ldots,\eta-1$}
\If{$c_i=0$}
\State $p_i\gets \mathcal{P}(U_\eta)p_i$
\EndIf
\EndFor
\State $p_\eta\gets \mathcal{P}(U_\eta)p_\eta$
\EndIf
\State \textbf{return} $\mathcal{A}(\phi_1\ldots\phi_\eta)$
\end{algorithmic}
\end{algorithm}

 Each ancilla qubit has an equal probability of being measured in either $\ket{0}$ or $\ket{1}$, and a phase correction is required only when an ancilla is measured in the $\ket{1}$ state for a qubit entangled with a swap. If the number of required phase corrections exceeds $\left\lfloor \frac{\eta}{2} \right\rfloor$, one can instead apply a minus sign to the terms that would not otherwise need correction, still yielding a correctly antisymmetrized wave function (up to an overall sign). In the worst-case scenario, this measurement-based approach reduces the number of applications of the state-preparation unitaries at each iterative step from $\eta-1$ to $\lfloor \eta/2\rfloor$ (not including the single application of $U_\eta$ that prepares the $\eta$-th particle in state $\phi_\eta$). On average, the total number of phase corrections for antisymmetrization of $\eta$ particles is given by
\begin{equation} 
 \langle N_\mathrm{corr}\rangle=   \sum_{n=2}^\eta \frac{1}{2^{n-1}}\sum_{j=0}^{n-1} \min(j,n-j)\left( \begin{array}{c}
         n-1  \\
         j 
    \end{array}\right),
\label{eq:av_num_corr}
\end{equation}
where $\langle N_\mathrm{corr}\rangle/(\eta(\eta-1)/2)\approx 0.5$ for large $\eta$. In Table \ref{tab:measurement_phase}, we explicitly show the phase application given any possible outcome of the ancilla measurements for systems of three and four particles.

\begin{table}[]
    \centering
    \caption{Phase corrections, based on the measurement outcome for three and four particles. In general, for antisymmetrizing the $\eta$-th particle with the antisymmetric state of $\eta-1$ particles, the maximum number of phase corrections will be $\lfloor \eta/2 \rfloor$.}
    \begin{tabular}{cc}
    \hline \hline
       Ancilla state  & Phase correction  \\
       \hline
       Three particles &  $\mathcal{P}(U_3)$ \\
        $\ket{00}$ &  $-$ \\
        $\ket{01}$ &   $p_1$ \\
        $\ket{10}$ &   $p_2$  \\
        $\ket{11}$ &  $p_3$ \\
               \hline
       Four particles &  $\mathcal{P}(U_4)$ \\
        $\ket{000}$ &  $-$ \\
        $\ket{001}$ &   $p_1$ \\
        $\ket{010}$ &   $p_2$  \\
        $\ket{100}$ &  $p_3$ \\
        $\ket{101}$ &  $p_1$ \& $p_3$ \\
        $\ket{110}$ &  $p_2$ \& $p_3$ \\
        $\ket{011}$ &  $p_1$ \&  $p_2$ \\
        $\ket{111}$ &  $p_4$ \\
        \hline \hline
     \end{tabular}
    \label{tab:measurement_phase}
\end{table}

\section{Complexity and comparison with other algorithms}
\label{sec:other_algorithms}

In this section, we study the scaling of our circuit and compare against other existing algorithms \cite{AbramsLloyd1997,Berry2018a,PhysRevA.106.032428}. We separate the discussion according to the nature of the states to be antisymmetrized: First, we consider the case of single basis states (integers, or positions in a lattice) in Sec. \ref{sec:integers}.  In Sec. \ref{sec:HFsd}, we extend the scaling study of our algorithm to more complicated orbitals, which are superpositions of single-particle basis states. 

\subsection{Antisymmetrization of single basis states}
\label{sec:integers}

Existing antisymmetrization algorithms \cite{AbramsLloyd1997,Berry2018a} assume that the input wave function is ordered
\begin{equation}
    \ket{r_1\ldots r_\eta},
\end{equation}
with $1\leq r_1<\ldots<r_\eta\leq N$ denoting the integers that label the first-quantized single-particle states. This condition is necessary in order to guarantee that the antisymmetrization operation
\begin{equation}
    \mathcal{A}(\ket{r_1\ldots r_\eta})=\frac{1}{\sqrt{\eta!}}\sum_{\sigma\in S_\eta}(-1)^{\pi(\sigma)}\ket{\sigma(r_1,\ldots,r_\eta)}
    \label{eq:antisym_mapping}
\end{equation}
is unitary and can therefore be implemented deterministically on a quantum computer. Here, $S_\eta$ is the symmetric group of order $\eta$, and $\pi$ is the parity of the permutation $\sigma$. Unlike our proposed method, these approaches rely on reversible sorting algorithms. Starting from a possibly disordered set of integers, these sorting algorithms produce the corresponding ordered set by performing a series of integer comparisons and swaps $(r_i,r_j)\rightarrow(\min(r_i,r_j),\max(r_i,r_j))$, storing the result of each comparison in an ancilla register. The stored ancilla values can later be used either to reverse the sorting procedure or, as the case may be, to apply the same series of permutations to an initially ordered set of integers.

The aforementioned algorithms begin the antisymmetrization procedure by introducing $\eta\lceil\log_2\eta\rceil$ additional qubit registers, wherein one prepares the symmetrized state on $\eta$ integers\footnote{In Ref. \cite{Berry2018a}, this state is obtained only after post-selection is used to remove terms with collisions (\textit{i.e.}, repeated integers). Moreover, this method prepares the symmetric state on $\eta$ integers chosen from $1,\ldots, f(\eta)$ where the requirement $f(\eta)>\eta^2$ guarantees greater than 50\% success probability, while $\eta\lceil\log_2f(\eta)\rceil$ qubits are needed.},
\begin{equation}
    |\Sigma_\eta\rangle=\frac{1}{\sqrt{\eta!}}\sum_{\sigma\in S_\eta}\ket{\sigma(1,\ldots,\eta)}.
\end{equation}
 One then performs a reversible sort on the symmetrized state $|\Sigma_\eta\rangle$, storing the result of each integer comparison in an ancilla register. These ancillae can then be used to map the ordered physical input state $\ket{r_1\ldots r_\eta}$, which has been separately prepared on $\eta k$ qubits, to its fully antisymmetrized form, as in Eq. \eqref{eq:antisym_mapping}, adding the appropriate phase factor $(-1)^{\pi(\sigma)}$ to each permutation. A final set of integer comparisons must then be performed on the particle registers in order to uncompute the ancilla registers. These integer comparisons are generally more computationally expensive than the integer comparisons used to sort the symmetric state; the particle integers have length $k=\lceil\log_2N\rceil$ whereas the ancilla registers have length $\lceil \log_2\eta\rceil$, and $N\gg \eta$ in cases where first quantization is advantageous. 
 
 The number of integer comparisons required to sort $\eta$ integers depends on the choice of sorting algorithm, which can be modeled as a network of successive comparisons and swaps, some of which can be performed in parallel. The most efficient sorting networks require $O(\eta\log \eta)$ comparators \cite{10.1145/800061.808726,10.1007/BF01840378,10.1145/2591796.2591830}; they achieve the theoretically optimal scaling for a sorting algorithm but have very large constant factors that render them practically useless for realistic values of $\eta$. The odd-even mergesort algorithm \cite{10.1145/1468075.1468121} requires $O(\eta\log^2\eta)$ comparators with a much more favorable constant factor, leading to $O(\eta\log^2 \eta \log N)$ gate complexity to perform antisymmetrization using the algorithm of Ref. \cite{Berry2018a}. 

 As mentioned above, the most expensive computational step in sorting-based antisymmetrization methods is the final set of comparisons --- of integers encoded on $k$ qubits --- required to uncompute the ancillae. Our algorithm uncomputes the ancilla qubits using an approach that does not rely on integer comparisons. To determine if two particles were previously swapped, we utilize the fact that the single-particle states $\phi_i$ are orthogonal. Because we construct the antisymmetric state on $\eta$ particles recursively, particle $i<\eta$ can only be in state $\phi_\eta$ if particles $i$ and $\eta$ were exchanged, thus allowing us to uncompute the corresponding ancilla qubit by controlling on the state of particle $i$. In the case that each $\phi_i$ corresponds to a single integer basis state $\ket{r_i}$, then the associated unitary $U_i$ is simply the product of Pauli $X$-gates that produce the binary representation of $r_i$. As a result, the leading computational costs in the non-measurement-based variant of our algorithm come from performing a total of $k \eta(\eta-1)/2$ \textsc{cswap} gates and $\eta(\eta-1)/2$ multi-controlled $C^k X$ gates, requiring a total of $O(\eta^2\log N)$ gates (note that each $C^k X$ gate can be performed with $O(k)$ single- or two-qubit gates, as detailed below). The measurement-based variant described in Sec. \ref{sec:improved} reduces these costs by a constant factor.
 
 In Table \ref{tab:algo_comp}, the asymptotic scaling behavior of our algorithm is compared to the antisymmetrization methods introduced in Refs. \cite{AbramsLloyd1997,Berry2018a}. Compared to Berry \textit{et al.}'s algorithm, the cost to antisymmetrize an initially ordered product state $\ket{r_1\ldots r_\eta}$ using our method scales the same ($\sim \log N$) with the number of states per particle but slightly worse with the number of particles: $\eta^2$ for our algorithm compared to $\eta\log^2 \eta$ for that of Berry \textit{et al}. \cite{Berry2018a}. 

\begin{table}[t]
    \centering
    \caption{Scaling comparison of $T$-gate complexity required to antisymmetrize a single ordered input product state $\ket{\Phi}=\ket{r_1\ldots r_\eta}$. The stated gate complexity of Berry \textit{et al.}'s algorithm assumes that the sorting network requires $O(\eta\log^2\eta)$ comparators, as is true of odd-even mergesort \cite{10.1145/1468075.1468121}. Note that our approach does not require the input state to be ordered.}
    \label{tab:algo_comp}
    \begin{tabular}{ll}
    \hline\hline
        Algorithm & ~~~~$T$-gate complexity \\ 
        \hline
       Abrams \& Lloyd \cite{AbramsLloyd1997} & ~~~~$O(\eta^2\log^2N)$ \\
       Berry \textit{et al.} \cite{Berry2018a} & ~~~~$O(\eta\log^2 \eta\log N)$ \\
       Present work & ~~~~$O(\eta^2\log N)$ \\
       \hline
       \hline
    \end{tabular}
\end{table}
 
The cost of sorting-based antisymmetrization methods is governed by the number of integer comparators required by the chosen sorting algorithm, with each comparator requiring $8k+O(1)$ $T$-gates \cite{Berry2018a}. Moreover, each comparator is associated to $k$ \textsc{cswap} gates, requiring $4k$ additional $T$-gates to implement, for a total cost of $12k+O(1)$ $T$-gates per integer comparison. The equivalent cost in our algorithm arises from the multi-controlled $C^k X$ gates required to uncompute the ancilla qubits, each with an associated \textsc{cswap} gate. Introducing $k-1$ additional ancilla qubits initialized in the $|0\rangle$ state, each $C^k X$ operation can be implemented with precisely $8(k-1)$ $T$-gates \cite{PhysRevA.87.042302}. Utilizing mid-circuit measurements and feed-forward, this cost can be further reduced to $4(k-1)$ $T$-gates per $C^k X$ operation \cite{PhysRevA.87.022328}. Combined with the cost of $k$ \textsc{cswap} gates, our algorithm requires $8k-4$ $T$-gates for each particle swap and uncomputation of the associated ancilla qubit.

Any exact implementation of the $C^kX$ gate requires at least $k$ $T$-gates \cite{Beverland_2020}. Alternately, one can implement a unitary that is within error $\epsilon_d$ in diamond distance of the desired $C^kX$ gate with at most $O(\log(1/\epsilon_d))$ $T$-gates \cite{Gosset:2025rjo}. The only non-Clifford gate required to construct this approximate unitary is a $C^jX$ gate where $j=\lceil\log_2(1/\epsilon_d)\rceil+3$. Utilizing this approach, our algorithm can be implemented with $4(k+j-1)$ $T$-gates for each particle swap and associated $C^kX$ gate. Clearly, this approach is only useful when $k>j$. While it could be advantageous for systems with very large numbers of single-particle states per particle, we will not consider this implementation further in this work.

 Assuming that odd-even mergesort is used and the number of sorted integers must be padded to a power of two, the number of comparators required to sort $\eta$ integers is precisely
 \begin{equation}
     N_\mathrm{comp}=2^{m-2}\left(m^2-m+4\right)-1,
 \end{equation}
 where $m=\lceil\log_2\eta\rceil$. By contrast, the number of \textsc{cswap} and multi-controlled $C^{k}X$ gates required by the non-measurement-based variant of our algorithm is
 \begin{equation}
     N_\mathrm{ctrl}=\frac{1}{2}\eta(\eta-1).
 \end{equation}
As discussed above, the leading $T$-gate cost of Berry \textit{et al}.'s algorithm scales as $12kN_\mathrm{comp}$ whereas our algorithm scales as $8kN_\mathrm{ctrl}$. We find that $12N_\mathrm{comp}\geq 8N_\mathrm{ctrl}$ for particle numbers in the ranges $2\leq\eta\leq14$, $17\leq\eta\leq 24$, $31\leq \eta\leq 40$, and $65\leq \eta \leq66$. Despite the nominal exponential scaling advantage of Berry \textit{et al.}'s algorithm with particle number, the need to pad the sorted array to a power of two for odd-even mergesort leads to particularly steep comparator overheads for particle numbers that barely exceed a power of two; for $\eta=64$ particles, $12N_\mathrm{comp}/8N_\mathrm{ctrl}\approx 0.40$ but for $\eta=65$ particles, $12N_\mathrm{comp}/8N_\mathrm{ctrl}\approx 1.06$.

\begin{figure}
    \centering
    \includegraphics[scale=0.7]{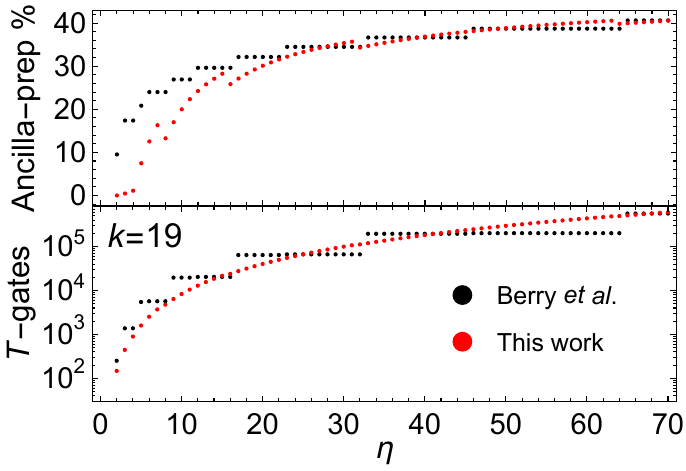}
    \caption{Bottom: Total number of $T$-gates required to antisymmetrize a system of $\eta$ particles with $k=19$ qubits per particle. The solid black points correspond to the method of Berry \textit{et al.} \cite{Berry2018a} with odd-even mergesort while the red points correspond to the non-measurement-based implementation of our algorithm. Top: The percentage of the total $T$-count associated with preparing the ancilla qubits (see text). We assume that the additional $O(1)$ $T$-gate cost per integer comparator required by Berry \textit{et al}.'s algorithm is zero.}
    \label{fig:Tcost_compare}
\end{figure}

The bottom panel of Fig. \ref{fig:Tcost_compare} shows a more precise comparison of the total $T$-gate count of our algorithm to that of Berry \textit{et al.} \cite{Berry2018a} for the purpose of antisymmetrizing single basis states (i.e., the single-particle states are integer labels). To compute the total $T$-count of their algorithm, we include the cost to reversibly sort $\eta$ registers of length $\lceil \log_2 f(\eta)\rceil=\lceil\log_2\eta^2\rceil$ as well as the cost to reversibly unsort $\eta$ registers of length $k=\lceil\log_2N\rceil$ using the odd-even mergesort network. Note that choosing $f(\eta)=\eta^2$ is the minimal requirement in order for the required post-selection to succeed with greater than 50\% probability. For $\eta>10$, the resulting success probability $P_s$ lies in the range $0.5<P_s<0.55$.

To compute the $T$-gate cost of our algorithm, we include the cost to prepare the required ancilla states as well as the cost to perform all requisite particle swaps and the ensuing uncomputation of all ancillae. As discussed in Appendix \ref{app:ancilla_prep}, preparing ancilla state $Y_n$ requires $2n-3$ single-qubit rotations. Preparing the antisymmetric state on $\eta$ particles, therefore requires a total of $\eta^2-4\eta+3$ arbitrary-angle single-qubit rotations.\footnote{When $n+1$ is a power of two, the state $Y_n$ can be prepared without the need for angle synthesis. Therefore the total angle count is less than the na\"ive count of $\eta^2-4\eta+3$ quoted above. Taking account of this effect when $\eta=50$, for example, the total ancilla preparation requires 2,204 rotations instead of 2,303. This more precise value is reflected in Fig. \ref{fig:Tcost_compare}.} Then, if the total operator norm error in the unitary that prepares all of the required $Y_n$ is $\epsilon_\mathrm{tot}$, the maximum error permitted in the synthesis of each single-qubit rotation is $\epsilon=\epsilon_\mathrm{tot}/(\eta^2-4\eta+3)$. This target error governs the $T$-count resulting from ancilla qubit preparation; in Fig. \ref{fig:Tcost_compare}, this cost is shown for $\epsilon_\mathrm{tot}=10^{-2}$ assuming that each single-qubit rotation is synthesized to error $\epsilon$ using the Ross-Selinger algorithm \cite{ross2014optimal}.

We consider a system with $k=19$ qubits per particle, which would be sufficient, for example, to represent each particle in a grid of size $64^3$ with two possible spin states. Due to the reduced cost of a single $C^kX$ gate --- $4(k-1)$ $T$-gates --- compared to the cost of a single integer comparison --- $8k+O(1)$ $T$-gates \cite{Berry2018a} --- our algorithm requires fewer total $T$-gates than does the sorting-based method of Berry \textit{et al.} for particle numbers $\eta\lesssim 40$, except in cases where $\eta$ is close to (without exceeding) a power of two, in which case odd-even mergesort does not require significant additional padding. (For purposes of comparison, we assume that the $O(1)$ $T$-gate cost of each integer comparison is zero.)

The top panel of Fig. \ref{fig:Tcost_compare} shows the fraction of the total $T$-gate cost associated with preparing the ancilla qubits in the required state. For the algorithm of Berry \textit{et al}., this is the cost to reversibly sort $\eta$ integers of size $\lceil\log_2 \eta^2\rceil$. In case of our algorithm, it is the cost to successively prepare the ancilla qubits in the states $Y_1,\ldots,Y_{\eta-1}$ to within total error $\epsilon_\mathrm{tot}$. The relative cost of our ancilla preparation dips when $\eta$ is a power of two, as the state $Y_{\eta-1}$ can be prepared exactly without arbitrary-angle rotations requiring synthesis, as described in Appendix \ref{app:ancilla_prep}.

The iterative nature of our algorithm suggests a hybrid approach: a sorting-based method can be used to prepare the antisymmetric state on an intermediate number of particles (\textit{e.g.}, the nearest power of two) with the resulting state supplied as input to our algorithm to recursively add the remaining particles. For example, the antisymmetric state on $\eta=65$ particles with $k=19$ qubits per particle could be constructed at a cost of 543 integer comparisons to construct the $64$-particle antisymmetric state followed by 64 \textsc{cswap} and $C^k$X gates to add the final particle using one recursive step of our algorithm for a total cost (including all necessary ancilla preparation) of $\approx 216,000$ $T$-gates. In contrast, 1471 integer comparators and $\approx 565,000$ $T$-gates are required to directly construct the $\eta=65$ antisymmetric state using the odd-even mergesort. This hybrid approach mitigates the $O(\eta^2)$ recursive overhead of our algorithm making it particularly efficient for large particle numbers.

Berry \textit{et al}. \cite{Berry2018a} also describe an antisymmetrization algorithm based on a quantum Fisher-Yates shuffle that --- like our algorithm --- does not rely on sorting and has gate complexity $O(\eta^2\log N)$. As in sorting-based approaches to antisymmetrization, the Fisher-Yates-inspired algorithm assumes that the input state is ordered.

\subsection{Orbital Slater determinants}
\label{sec:HFsd}

The antisymmetrization algorithms introduced in Refs. \cite{AbramsLloyd1997,Berry2018a} are defined to act on a single \textit{ordered} integer basis state $\ket{r_1\ldots r_\eta}$, with the understanding that the same method applies to all linear superpositions of such states. That is, each integer basis state in the superposition must be ordered. However, consider the case of particles occupying Hartree-Fock (HF) orbitals, where each orbital is a linear superposition of the single-particle integer basis states $\ket{\phi_i}=\sum_{j}c_{i,j}\ket{r_j}$, and it is assumed that the HF orbitals are orthonormal, $\langle\phi_{i_1}|\phi_{i_2}\rangle=\delta_{i_1,i_2}$. In such case, the HF state $\ket{\phi_1\ldots\phi_\eta}$ is a superposition of many integer basis states $\ket{r_{j_1}\ldots r_{j_\eta}}$, which are not ordered a priori.

\begin{table}[t]
    \centering
    \caption{Scaling comparison of $T$-gate complexity required to antisymmetrize a product state of non-trivial single-particle orbitals $\ket{\Phi}=\ket{\phi_1\ldots \phi_\eta}$.}
    \label{tab:algo_comp_orbitals}
    \begin{tabular}{ll}
    \hline\hline
        Algorithm & ~~~~$T$-gate complexity \\ 
        \hline
       Babbush \textit{et al.} \cite{Babbush:2023sas} & ~~~~$O(\eta N)$ \\
       Present work & ~~~~$O(\eta^2 \sqrt{N})$ \\
       \hline
       \hline
    \end{tabular}
\end{table}

Assuming that the HF orbitals are sufficiently far from trivial, the cost of antisymmetrization will be subdominant to the cost of state preparation. The superposition of \textit{ordered} integer basis states required by existing antisymmetrization algorithms can be prepared with $O(\eta N\log N)$ gate complexity \cite{Babbush:2023sas}, including $O(\eta N)$ Toffoli gates. Our algorithm prepares the $\phi_i$ states independently on each single-particle Hilbert space. An arbitrary dimension-$N$ quantum state can be prepared up to error $\varepsilon$ by a Clifford$+T$ circuit with $O(\log(N/\varepsilon))$ clean ancillas, a tunable number of $\sim(\lambda\log(\frac{\log N}{\varepsilon}))$ dirty ancillas, and $O\left(\frac{N}{\lambda}+\lambda \log\frac{N}{\varepsilon}\log\frac{\log N}{\varepsilon}\right)$ $T$-gates \cite{Low2024tradingtgatesdirty}. With some algorithmic improvements \cite{Gosset:2024vku}, the theoretically optimal $T$-gate scaling can be achieved by a quantum circuit with $O\left(\sqrt{N\log(1/\varepsilon)}+\log(1/\varepsilon)\right)$ $T$-gates and ancillas. 

The number of qubits required to achieve $O(\sqrt{N})$ $T$-gate complexity grows faster than logarithmic in the number of single-particle states, but slower than any polynomial. In particular, it is still more favorable than the linear qubit requirements of second quantization. As the $O(\sqrt{N})$ scaling corresponds to dirty ancillas, while acting only on the $k=\lceil\log N\rceil$ qubits of a single particle's Hilbert space, one can in principle utilize the remaining $(\eta-1)k$ qubits of the other particles to mitigate somewhat the ancilla requirements. Still, depending on hardware limitations and the nature of the system under consideration, $O(\sqrt{N})$ ancilla scaling may not be practical. In such case, single-particle states can instead be prepared with $O(\log N)$ ancillas and $O(N)$ $T$-gates \cite{Low2024tradingtgatesdirty}. 

Assuming that one can employ $O(\sqrt{N})$ ancillas, each $U_i$ (equivalently, $U_i^\dagger)$ operator can be (approximately) implemented with $O(\sqrt{N})$ $T$-gates \cite{Gosset:2024vku}. To prepare the antisymmetric state on $\eta$ particles using our method requires $O(\eta^2)$ total applications of either $U_i$ or $U_i^\dagger$ operators, leading to a total $T$-gate cost $O(\eta^2\sqrt{N})$. Thus, for the purpose of preparing an antisymmetric state of non-trivial single-particle states, our algorithm exhibits superior $T$-gate scaling compared to the best known alternatives whenever $\eta<\sqrt{N}$, which should be true in many scenarios where first quantization is advantageous over second quantization. These results are summarized in Table \ref{tab:algo_comp_orbitals}.

The sorting-based antisymmetrization algorithm described by Berry \textit{et al.} requires only natively fault-tolerant gates; that is, there are no arbitrary-angle rotations requiring synthesis in terms of, say, Clifford$+T$ gates. Of course, if the single-particle states to be antisymmetrized are non-trivial superpositions of the integer basis states, then arbitrary-angle rotations will generically be required in state preparation (specifically in the form of Givens rotations in Refs. \cite{PhysRevA.106.032428,Babbush:2023sas,PRXQuantum.6.020319}). Each variation of our algorithm employs $R_y$ gates with non-trivial angles in order to prepare the ancillae in the requisite initial states. On the other hand, Berry \textit{et al.}'s algorithm is non-deterministic, requiring measurements in order to project onto the collision-free subspace, though the authors provide a prescription to guarantee that the success probability is greater than $50\%$.  

 As we have emphasized, our approach does not assume that the input state is ordered in terms of the integer basis states; indeed the $\phi_n$ states that we prepare may be complicated superpositions of the integer basis states and therefore have no intrinsic sense of ordering. Rather, our input state is ordered in the sense that the state $\phi_n$ is initially prepared in the registers of particle $n$, and due to the recursive nature of our algorithm each particle register is treated uniquely. For this reason, our algorithm is reversible and, when reversed, uniquely maps the fully antisymmetrized state $\mathcal{A}(\phi_1\ldots\phi_\eta)$ to a particular initial ordering.

 As in the previous section, we can again envision a hybrid approach: a sorting-based method (or alternative antisymmetrization approach e.g., \cite{10.1063/5.0239980}) can be used to prepare the antisymmetric state on an intermediate number of particles with the resulting state supplied as input to our algorithm to recursively add the remaining particles. For example, if an antisymmetric state is prepared for a ``target'' consisting of many fermions, an additional ``projectile'' consisting of a single particle can be added to create a totally antisymmetric state using a single step of our algorithm. Alternately, if both projectile and target are composite systems, antisymmetric states can be prepared separately for each cluster before antisymmetrizing across clusters \cite{Stetcu:2025bbu}, allowing one to exploit additional efficiencies with respect to state preparation and uncomputation of ancilla qubits.

\section{Fault-tolerant implementation}
\label{sec:examples}

At present, most approaches to quantum error correction are not directly applicable to arbitrary-angle single-qubit rotations, due to error-correcting codes being discrete and rotations being continuous  \cite{NielsenMichaelA.2000Qcaq}. The few existing efforts to implement continuous-variable error correction have so far been limited in scope to small rotation angles \cite{choi2023fault, mayer2024benchmarking,luthra2025unlocking, he2025high, sethi2025rescq}. Thus, to produce a fault-tolerant quantum circuit, we express all arbitrary-angle rotations in terms of a discrete set of Clifford+$T$ gates using the Ross-Selinger algorithm, which is provably optimal \cite{ross2014optimal}. While the Ross-Selinger algorithm  decomposes $Z$-rotations, changes-of-basis to the $X$ and $Y$ bases are trivial with additional Clifford gates. The only multi-qubit gate native to this gate set is the \textsc{cnot} gate; decompositions of other relevant multi-qubit gates into the Clifford+$T$ gate set can be found in Appendix \ref{app:circuitelementdecomp}.

The impact of hardware noise is evaluated using the test case of an $\eta=3$ particle system with $k=3$ qubits representing the state of each particle. We perform our noise study using the measurement-based Algorithm \ref{algo:measure_antisym}. Due to its lower gate overhead,  we expect Algorithm \ref{algo:measure_antisym} to be less sensitive to hardware noise than Algorithm \ref{algo:recur_antisym}. In our test example, each fermion can be in one of three basis states that encode integers $0$, $1$ and $2$, and thus the single-particle states can be simply prepared with Pauli $X$-gates. The nontrivial part of the implementation lies in the creation of the $Y_2$ ancilla state, which requires one single-qubit rotation about the $Y$-axis with angle $\theta=2 \arccos{\sqrt{\frac{1}{3}}}$. The Ross-Selinger algorithm was used to generate discrete approximations of a $Z$-axis rotation with the same angle, which was then basis-transformed into a $Y$-axis rotation. The approximation can be improved to arbitrary accuracy $\varepsilon$ --- defined as the operator norm of the difference between the exact and approximate rotation operators --- at the cost of additional circuit depth $\sim \log(1/\varepsilon)$. 

The gate counts required to represent $R_y(2 \arccos{\sqrt{\frac{1}{3}}})$ in terms of Clifford+$T$ gates up to error $\varepsilon$ are given in Table \ref{tab:rs_error_data} for $\varepsilon$ ranging from $0.1$ to $10^{-13}$. Not including gate costs associated with rotation synthesis or phase corrections, the circuit for our $(\eta=3,k=3)$ example contains 108 Clifford gates and 65 $T$-gates. Each phase correction $\mathcal{P}(U)$ requires an additional 10 Clifford gates (not including the $X$-gate needed to uncompute the measured ancilla if it needs to be reused) and 7 $T$-gates, the latter arising from the multi-controlled phase gate. (For comparison, Algorithm \ref{algo:recur_antisym} requires 171 Clifford gates and 110 $T$-gates, not including gates arising from rotation synthesis.)

\begin{table}[h]
    \centering
    \caption{The number of $T$-gates and total gates required to approximate $R_y(2 \arccos{\sqrt{\frac{1}{3}}})$ in Clifford+$T$ gates to various accuracies using the Ross-Selinger algorithm. The stated error ($\varepsilon$) represents the operator norm of the difference between the exact and approximate rotation operators. The state infidelity resulting from rotation synthesis is roughly the square of the difference operator norm.}
    \label{tab:rs_error_data}
    \begin{tabular}{ccc}
        \hline\hline
        Error & $T$-gate count & Total gate count \\
        \hline
        $1 \times 10^{-1}$ & 8 & 28 \\
        $9 \times 10^{-3}$ & 22 & 64 \\
        $1 \times 10^{-3}$ & 34 & 91 \\
        $8 \times 10^{-6}$ & 60 & 158 \\
        $1 \times 10^{-7}$ & 82 & 215 \\
        $7 \times 10^{-11}$ & 130 & 340 \\
        $1 \times 10^{-13}$ & 168 & 432 \\
        \hline\hline
    \end{tabular}
\end{table}

We simulate noise in Qiskit \cite{Javadi-Abhari:2024kbf} through a model consisting of a depolarizing channel where each Clifford gate in the circuit has the same infidelity and each $T$/$T^{\dag}$ gate has the same infidelity. The infidelities are adopted from recent examples in the literature, including a best-case (drawn from 2024-2025 results for trapped-ion logical qubits), a moderate-case (drawn from 2024-2025 results for neutral-atom logical qubits), and a worst-case (drawn from 2021 results for trapped-ion logical qubits). For Clifford gates, the infidelities are $5 \times 10^{-6}$ \cite{paetznick2024demonstration}, $9 \times 10^{-4}$ \cite{bluvstein2024logical}, and $3 \times 10^{-3}$ \cite{egan2021fault}, respectively. For $T$ and $T^{\dag}$ gates, the infidelities are $5 \times 10^{-4}$ \cite{daguerre2025experimental}, $6 \times 10^{-3}$ \cite{sales2025experimental}, and $2 \times 10^{-2}$ \cite{egan2021fault}. The depolarizing parameter for 1-qubit gates is obtained by multiplying the desired infidelity by 2, and the depolarizing parameter for 2-qubit gates is obtained by multiplying the desired infidelity by $\frac{4}{3}$.

The effect of the noise in these test cases is gauged using two methods. The first is state fidelity, which is defined between two density matrices $\rho$ and $\sigma$ like so \cite{Jozsa01121994}:

\begin{equation}
    F(\rho, \sigma) = \left( \mathrm{tr} \sqrt{\sqrt{\sigma} \rho \sqrt{\sigma}} \right)^2.
    \label{eq:fidelitydef}
\end{equation}

\noindent The density matrices for this calculation are obtained from the particle-qubit registers at the end of the antisymmetrization circuit. The $\sigma$ matrix is drawn from a noise-free sample, while the $\rho$ matrix is drawn from the noisy sample. 

\begin{figure}
    \centering
\includegraphics[scale=0.84]{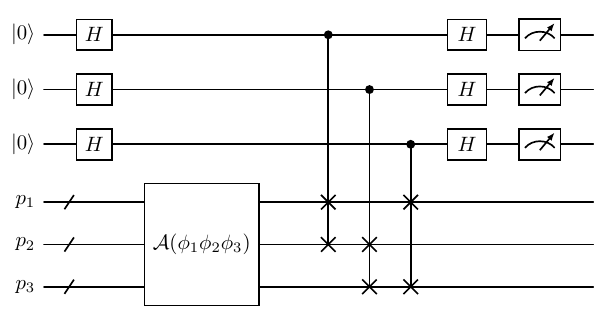}
    \caption{Antisymmetry probability circuit for a system of three particles.}
    \label{fig:antisym_circuit}
\end{figure}

The second metric we use to evaluate the effect of noise on the prepared state is the antisymmetric state probability, which is found by appending the circuit element shown in Fig. \ref{fig:antisym_circuit} to the end of the state-generation circuit and then measuring the ancillae. If the prepared state is antisymmetric with respect to a given two-particle swap, then the corresponding ancilla qubit should be purely in the $\ket{1}$ state after the circuit in Fig. \ref{fig:antisym_circuit} is executed. Thus, after measuring the 3 ancilla qubits, the probability of obtaining the $\ket{111}$ state corresponds to the antisymmetry probability.

\begin{figure*}
    \centering
    \includegraphics[scale=0.5]{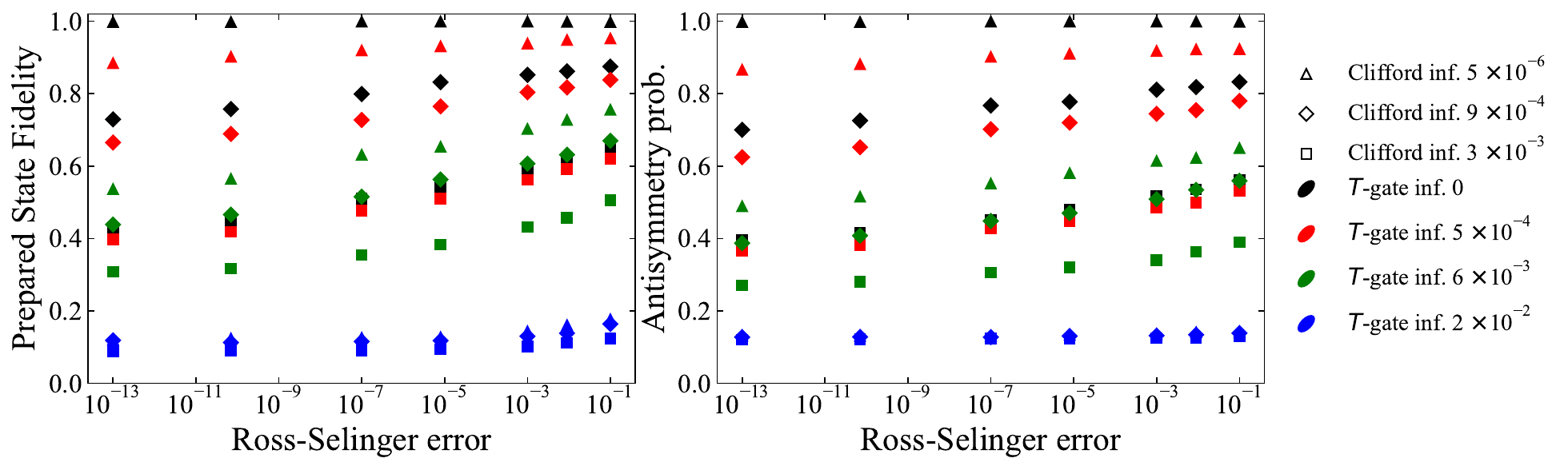}
    \caption{Fidelity (left) and antisymmetric probability (right) of the 3-particle state produced by the measurement-based variant of our algorithm, as a function of the Ross-Selinger error. Infidelities of Clifford and $T$-gates under a depolarizing noise channel are indicated by shape and color, respectively (see plot legend).}
    \label{fig:antisym_noise_results}
\end{figure*}

Figure \ref{fig:antisym_noise_results} illustrates the effect of noise on the prepared state for different gate infidelities and varying degrees of accuracy in the rotation-gate synthesis. Excluding the most optimistic case ($T$-gate infidelity zero, Clifford-gate infidelity $5 \times 10^{-6}$), we find that the circuit with the largest Ross-Selinger error ($\varepsilon=10^{-1}$) produces the highest circuit fidelity. As shown in Table \ref{tab:rs_error_data}, more accurate rotation synthesis requires larger gate counts; for the particular circuits and gate fidelities that we consider, the noise from these additional gates outweighs the marginal increase in the accuracy of the approximate rotation gate, ultimately leading to lower fidelity in the prepared state. Thus, for near-term applications it is preferable to limit the depth of the rotation synthesis.  

For Clifford and $T$-gate infidelities of comparable magnitude, the final state's fidelity is more sensitive to the Clifford infidelities. However, the higher magnitude $T$-gate errors in state-of-the-art hardware means that currently the $T$-gate error remains the limiting factor. Finally, the antisymmetric state probability is correlated with the fidelity, although the former is consistently lower than the latter due to the additional noise in the ancilla-related operations. For circuits that are too large for direct sampling of the final-state density matrix to be feasible, the antisymmetric state probability measurement circuit from Fig. \ref{fig:antisym_circuit} can serve as an effective proxy for state-fidelity results. 

\section{Summary and conclusions}
\label{sec:conclusions}

We presented a novel recursive algorithm for the antisymmetrization of an arbitrary number of identical fermions. When preparing antisymmetric states of simple integer labels, our algorithm generally exhibits a lower total $T$-gate count than existing sorting-based algorithms for particle numbers $\eta\lesssim 40$ except when $\eta$ is close to (without exceeding) a power of two. When constructing antisymmetric states of complicated single-particle orbitals, we demonstrated that each $U_i$ (or $U_i^\dagger$) operator can be implemented with $O(\sqrt{N})$ $T$-gates, leading to a total cost of $O(\eta^2 \sqrt{N})$ for antisymmetric state preparation. This scaling outperforms the best known alternatives whenever $\eta < \sqrt{N}$, a condition expected to hold in many regimes where first quantization offers clear advantages. The qubit requirements for achieving this $T$-gate complexity grow faster than logarithmic in the number of single-particle states but remain more favorable than the linear scaling of second quantization. Furthermore, localization of the trial wave function can be exploited to limit the effective number of active qubits for state preparation, thus reducing the number of ancillae. Combined with recent progress in simulating nuclear interactions in first quantization \cite{Weiss:2024mie}, this work has the potential to enable simulations of few-body nuclear systems using quantum computers in the near future. 

A detailed noise analysis of our algorithm was performed on a example 3-particle system. To ensure compatibility with state-of-the-art quantum error correction methods, which typically do not handle arbitrary rotation gates efficiently, the non-Clifford rotation required by our algorithm was compiled into the Clifford+$T$ gate set using the Ross–Selinger algorithm. The required single-qubit rotation was synthesized to varying degrees of accuracy, with smaller errors requiring larger gate counts.

The results of our noise study --- performed with the measurement-based variant of our algorithm due to its lower gate overhead --- highlight three key findings: First, except in the noiseless limit and when Clifford errors dominate, the Ross-Selinger approximation with error $\varepsilon=10^{-1}$ yields the highest circuit fidelity, showing that near-term noise has a stronger impact than synthesis inaccuracy. Second, although fidelity is more sensitive to Clifford errors when gate infidelities are comparable, the typically larger $T$-gate errors in current hardware make them the primary limitation. Finally, the antisymmetric state probability, while consistently below fidelity, tracks it closely, indicating its utility as a scalable proxy for fidelity in circuits too large for full density-matrix sampling. Together, these findings point to practical strategies for optimizing near-term quantum computations under realistic noise conditions.

\section*{Acknowledgments}

 The authors thank M. Anghel and A. Baroni for stimulating discussions about antisymmetrization in the first quantization, and L. Cincio for help with the Ross-Selinger algorithm. This work was carried out under the auspices of the National Nuclear Security Administration of the U.S. Department of Energy at Los Alamos National Laboratory under Contract No. 89233218CNA000001. ER is supported by the National Science Foundation under cooperative agreement 2020275 and by the U.S. Department of Energy through the Los Alamos National Laboratory. IC, IS and JC gratefully acknowledge support by the Advanced Simulation and Computing (ASC) program. IC and JC also acknowledge the Quantum Science Center for partial support of their work on this project. RW acknowledges support by the Edwin Thompson Jaynes Postdoctoral Fellowship of the Washington University Physics Department. ChatGPT and Scopus AI were used to aid in literature searches for background material.

\appendix
\section{Preparation of ancillae}
\label{app:ancilla_prep}
In each variant of our antisymmetrization algorithm, we introduce $\eta-1$ ancilla qubits that must be prepared in the state
\begin{equation}
\begin{split}
    \ket{Y_{\eta-1}}&\equiv \frac{1}{\sqrt{\eta}}\left(1-\sum_{j=1}^{\eta-1}X_j\right)\ket{0}^{\otimes (\eta-1)}\\
    &=\frac{1}{\sqrt{\eta}}\left(\ket{0}^{\otimes (\eta-1)}-\sqrt{\eta-1}|W_{\eta-1}\rangle\right),
    \label{eq:YN}
    \end{split}
\end{equation}
where the $W_{\eta-1}$ state is well studied as an archetype of multi-particle entanglement \cite{Dur:2000zz}. For generic $n$, we can prepare the state $Y_n$ by adapting the linear-time-complexity $W_n$ state-preparation algorithm presented in Ref. \cite{Cruz_2019}.

\begin{figure}
    \centering
    \includegraphics[scale=1.0]{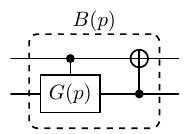}
      \caption{Fundamental building block $B(p)$ consisting of a controlled-$G(p)$ rotation followed by an inverted \textsc{cnot}.}
    \label{fig:Bp}
\end{figure}

For $0<p<1$, we define the gate
\begin{equation}
    G(p)\equiv \left(\begin{array}{cc}
       \sqrt{p}  & -\sqrt{1-p} \\
       \sqrt{1-p}  & \sqrt{p}
    \end{array}\right),
\end{equation}
which is equivalent to $R_y(\theta)$ with angle satisfying $\cos(\theta/2)=\sqrt{p}$. The fundamental building block $B(p)$ is a controlled-$G(p)$ rotation followed by an inverted \textsc{cnot}, as depicted in Fig. \ref{fig:Bp}.

 The $W_n$ state is prepared by initializing the first qubit in state $|1\rangle$ and all other qubits in state $\ket{0}$. The desired equal superposition is then prepared by acting the sequence of building block gates
\begin{equation}
    B(1/n),B(1/(n-1)),\ldots, B(1/3),B(1/2)
    \label{eq:Bp_sequence}
\end{equation}
on pairs of adjacent qubits $(q_1,q_2), (q_2,q_3),\ldots, (q_{n-1}, q_n)$. 

\begin{figure*}
    \centering
    \includegraphics[scale=1.0]{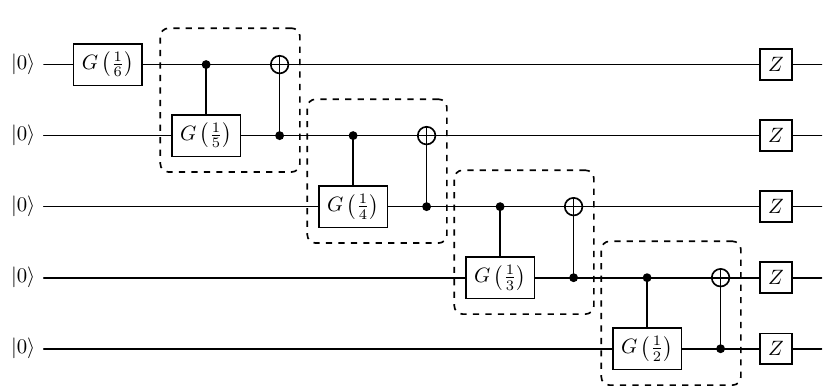}
    \caption{Quantum circuit that prepares the state $Y_5$.}
    \label{fig:Y5_prep}
\end{figure*}

To prepare instead the $Y_n$ state, we begin with all qubits in the $\ket{0}$ state and act the rotation $G(1/(n+1))$ on the first qubit. Then, we act the same sequence of $B(p)$ gates in Eq. \eqref{eq:Bp_sequence}. Finally, we produce the desired relative signs by acting Pauli $Z$ on each qubit. An example quantum circuit that prepares the state $Y_5$ is shown in Fig. \ref{fig:Y5_prep}. Preparing the state $Y_n$ using this method requires 1 rotation and $n-1$ controlled rotations. The controlled rotation $G(\frac{1}{2})$ can be implemented via a controlled Hadamard. The remaining $n-2$ controlled rotations can be decomposed into circuits each containing two \textsc{cnot} gates and two single-qubit half-angle rotations. Therefore, constructing the state $Y_n$ requires a total of $2n-3$ arbitrary-angle rotations.

\begin{figure*}
    \centering
    \includegraphics[scale=0.9]{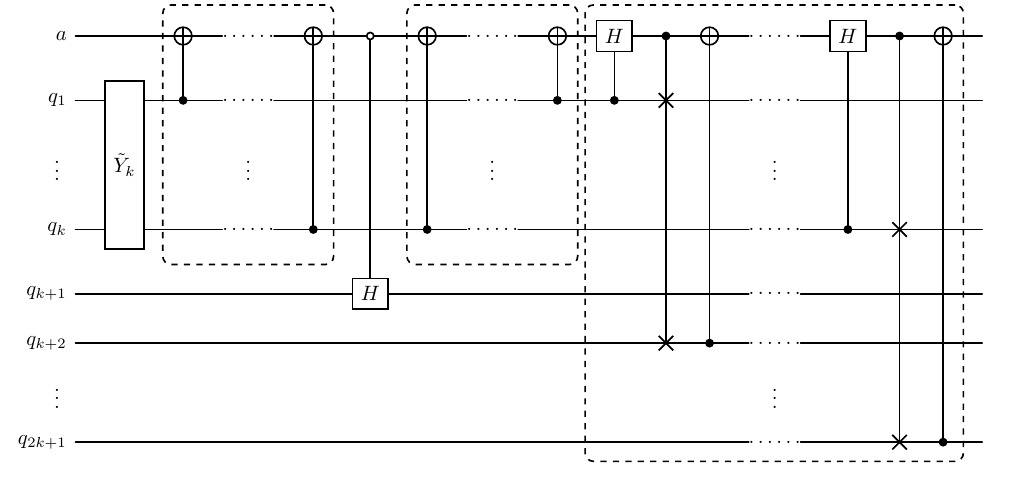}
    \caption{Quantum circuit that prepares the state $\tilde{Y}_{2k+1}$ on $2k+1$ qubits starting from the state $\tilde{Y}_k$ prepared on $k$ qubits. It is assumed that $k+1$ is a power of two.}
    \label{fig:YN_prep_power2}
\end{figure*}

When $n+1$ is a power of two, we can use an alternate approach to prepare the state $Y_n$ without arbitrary-angle rotations. Let $\tilde{Y}_n$ denote the state in Eq. \eqref{eq:YN} without the relative minus signs (\textit{i.e.}, $|\tilde{Y}_n\rangle=\prod_{i=1}^{n-1}Z_i \ket{Y_{n}}$). When $n+1$ is a power of two, the state $\tilde{Y}_n$ can be prepared recursively using a single ancilla as shown in Fig. \ref{fig:YN_prep_power2}. The base case is $\tilde{Y}_1=H\ket{0}$.

\newpage

\section{Decomposition of multi-qubit circuit elements}
\label{app:circuitelementdecomp}

\begin{figure}[h]
    \centering
\includegraphics[scale=0.75]{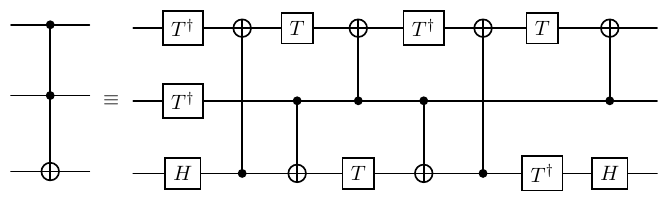}
    \caption{Decomposition of the double-controlled X (Toffoli) gate into native members of the Clifford$+T$ gate set, using the method in Fig. 7(a) of Ref. \cite{amy2013meet}. The CCZ gate can be represented by the same decomposition with the $H$ gates on the target qubit removed.}
\label{fig:ccx_decomposition}
\end{figure}

\begin{figure}[h]
    \centering

\includegraphics[scale=0.7]{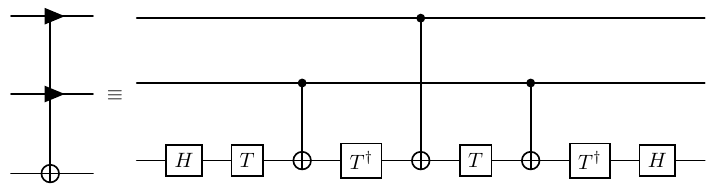}
    \caption{Decomposition of the double-controlled X (Toffoli) gate up to a phase into native members of the Clifford$+T$ gate set using the method in Ref. \cite{Weiss:2024mie}.}
\label{fig:phase_ccx_decomposition}
\end{figure}

\begin{figure}[h]
    \centering
\includegraphics[scale=0.6]{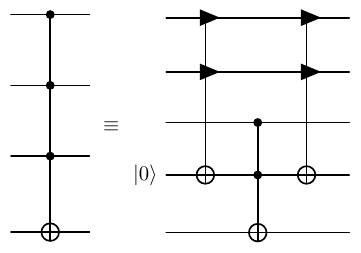}
    \caption{Decomposition of the multi-controlled X gate into native members of the Clifford$+T$ gate-set that provides a method of implementing an arbitrary number of control qubits $n_c$ using $n_c - 2$ ancilla qubits, based on a similar method in Refs. \cite{PhysRevA.87.042302,Weiss:2024mie}. This case is for $n_c =3$. }
\label{fig:mcx_decomposition}
\end{figure}

In Sec. \ref{sec:examples}, a simulated device with a native gate set composed of $T$, $T^{\dagger}$, all one-qubit Clifford, and \textsc{cnot} gates is used as a proxy for real-world noisy devices. In addition to these gates, the circuits that implement the algorithms introduced in Sec. \ref{sec:algorithm} require $Y$-rotations, controlled $Y$-rotations, controlled \textsc{swap} gates, and multi-controlled $X$ and $Z$ gates. As discussed in Sec. \ref{sec:examples}, the $Y$-rotations can be synthesized into Clifford+$T$ gates using the Ross-Selinger algorithm. The controlled-$Y$ rotations are decomposed into \textsc{cnot} gates and $Y$-rotations (the latter of which is again addressed by the Ross-Selinger approximation). In the special case of the controlled $G(\frac{1}{2})$ operation, this gate can be decomposed into a controlled-$H$ gate followed by a \textsc{cnot} gate, enabling an exact expression in terms of the native gate-set. The controlled-$H$ can in turn be represented with 1 \textsc{cnot} gate surrounded by one $T$ and one $T^{\dagger}$ gate and a basis-transformation on the target qubit that applies the following map to the Pauli gate-set: $\{\sigma^{(x)} \rightarrow \sigma^{(z)},  \sigma^{(y)} \rightarrow -\sigma^{(x)}, \sigma^{(z)} \rightarrow -\sigma^{(y)} \}$. \textsc{swap} gates are built out of 3 \textsc{cnot} gates that alternate in direction, so a controlled \textsc{swap} gate can be built by replacing the middle \textsc{cnot} with a Toffoli gate. Figures \ref{fig:ccx_decomposition}--\ref{fig:mcx_decomposition} in this Appendix detail how to express the remaining non-native gates, namely the multi-controlled X and Z gates, in terms of the native gate-set.

\newpage

\newpage

\bibliographystyle{quantum}
\bibliography{references}

\end{document}